\documentclass[11pt]{article}
\usepackage[a4paper, total={16cm, 23cm}]{geometry}
\usepackage{amsmath}
\usepackage{graphicx,psfrag,epsf}
\usepackage{enumerate}
\usepackage{enumitem}   
\usepackage{caption}
\usepackage{amsfonts}
\usepackage{longtable}
\usepackage{placeins}
\usepackage{subcaption}
\usepackage[utf8]{inputenc}
\usepackage[english]{babel}
\usepackage{blkarray}
\usepackage[hidelinks,colorlinks=true,allcolors=blue]{hyperref}

\usepackage{mathrsfs}
\usepackage{url}
\usepackage{verbatim}
\usepackage{listings}
\usepackage{tikz}
\usepackage{tcolorbox}
\usetikzlibrary{shapes.geometric, arrows}
\usepackage[natbibapa]{apacite}
\newcommand\blfootnote[1]{%
  \begingroup
  \renewcommand\thefootnote{}\footnote{#1}%
  \addtocounter{footnote}{-1}%
  \endgroup
}
\usepackage{pdflscape}

\title{Capturing Perception to Poverty using Conjoint Analysis \& Partial Profile Choice Experiment}
\author{Anushka De , Diganta Mukherjee\\ Indian Statistical Institute, Kolkata}
\date{}

\begin{document}

\maketitle

\begin{abstract}
The objective of this study is applying a utility based analysis to a comparatively efficient design experiment which can capture people's perception towards the various components of a commodity. Here we studied the multi-dimensional poverty index and the relative importance of its components and their  two-factor interaction effects. We also discussed how to model a choice based conjoint data for determining the utility of the components and their interactions. 
Empirical results from survey data shows the nature of  coefficients, in terms of utility derived by the individuals, their statistical significance and validity in the present framework. There has been some discrepancies in the results between the bootstrap model and the original model, which can be understood by surveying more people, and ensuring comparative homogeneity in the data. 
\end{abstract}

\section{Introduction}
\blfootnote{Abbreviations used: AIC: Akaike information criterion, CM: Conjoint Measurement, MNL: Multinomial Logit model, RUT: Random Utility Theory}


A common issue faced in the literature of Economics, Marketing, Statistics, Psychology and other related field is in capturing consumer perception towards a particular commodity. It is often possible to decompose the various components of a product (e.g. the brand, screen size, visual quality of a television). In the context of Multidimensional Poverty Index, (\cite{OPHI}), we may think of Poverty as a composite commodity consisting of attributes like Nutrition, Child Mortality, Years of Schooling, School Attendance, Cooking Fuel Sanitation, Drinking Water, Electricity, Housing, Assets. The MPI already assigns equal weights to Health, Education and Living Standards. 
Our problem of interest is to suitable capture the perception of people towards the various dimensions of the MPI, re-estimating the weights of the attributes. We also aim to describe the nature of association, if any, between any two attributes. 
\cite{alkire2010a_robustweights} demonstrates that the Multidimensional Poverty Index, developed by OPHI for the UNDP Human Development Report, is robust to a range of different weights. However there is sufficient gap in the literature in methods to re-estimating the weights with respect to consumer preferences. Furthermore, there is no work to assess the inter-relationship between any two attributes of the MPI. \\
\cite{PaulGreen_Rao} introduced the method of Conjoint analysis to estimate the effects of a set of independent variables in such problems. Conjoint Analysis is a generic term used to describe  any decompositional method that estimates the structure of a consumer’s preferences in terms of the levels of attributes of the alternatives. The methodology quite
heavily uses statistical experimental design and parameter estimation methods. Conjoint analysis is quite closely related to other developments in Information Integration Theory and its associated method of Functional Measurement (\cite{Anderson}). 

In contrast to traditional Conjoint Analysis that relies on Conjoint Measurement (CM), which is not a behavioral theory (of choice), Discrete Choice Experiments are based on a long-standing, well-tested theory of choice behavior that can take inter-linked behaviors into account. The theory was proposed by \cite{Thurstone}, and is called random utility theory (RUT). \cite{LOUVIERE201057} compares the two methods: Conjoint Analysis and Discrete Choice Experiments.
\par In a similar context we consider generic stated choice experiments in which the choice sets are pairs of alternatives and where one is interested in estimating the main effects and two-factor interactions of a potentially large number (say ten or more) of qualitative attributes. Every alternative is represented by a combination of attribute levels and in the marketing literature these combinations are usually referred to as ``profiles''. When all attributes in the experiment are used to specify the alternatives, they are called full profiles. By contrast, alternatives which are specified by using only a subset of the alternatives are called partial profiles. Considering the practical aspect in a survey experiment, partial profiles are more preferred. \cite{GROMANN2019100136} presents a practical approach to designing statistically efficient partial-profile choice
experiments with two alternatives for estimating main effects and
interactions of many two-level attributes. The performance of the design is assessed by the D-efficiency criterion. This amounts to calculating the
suitably normalized ratio of the determinant of the Fisher information matrix to a corresponding benchmark value which is usually
derived mathematically. With this criterion, one seeks to find designs for which the determinant of the Fisher information matrix is large
or, equivalently, the determinant of the asymptotic covariance matrix of the maximum-likelihood estimator is small.

\par The objective of this study lies to re-estimate the weights of Multi-dimensional Poverty index and determining the nature of relation between any two attributes (say, Nutrition and Maternal Health, Nutrition and Housing etc). 
This paper consists of analysing a choice based data collected using the method of  \cite{GROMANN2019100136} and thereafter instead of using the assumptions of Mutinomial Logit model, a utility based analysis of the consumer choices has been adapted. 
\par The remainder of the paper is organized as follows. Section 2 describes the mathematical model incorporated and the method of estimation of individual weights of the attributes and their 2-factor interaction. Section 3 presents an application where such a design scheme and model has been applied to estimate the relative weights of the components of Multi-dimensional poverty index, their significance and the nature of the two-factor interaction. Section 4 concludes the paper. All tables related to the analysis are collected in the Appendix.

\section{Modelling Utility : Multinomial Logit Model for Choice Based Conjoint Data}

We use the choice model, called the multinomial logit model as described in \textit{ \cite{Rao}} for analysing choice based conjoint data. This model is highly versatile and is based on sound theoretical assumptions.
As a first step we recognize that the utility of an
item consists of two components: a deterministic component and a random component.
The deterministic component can be modeled in terms of various factors. The
random component is assumed to represent all unobserved factors in the choice
process. 

Denote by $u_j$ the utility derived by the individual from profile $j$ of the choice set $S$. It can be decomposed as: 
$u_j=v_j+\epsilon_j$ where $v_j$ is the deterministic component and $\epsilon_j$ is the random component of the utility. \\

{\bf Modelling the Deterministic Component:}
Based on the premise that the deterministic component of a utility is derived from the attributes (here, components of multidimensional poverty index), we can model $v_j$ as follows: \begin{align}
    v_j=\sum_{r=1}^p \beta_{r}x_{rj}+\sum_{r=1}^p\sum_{r'=1,r\neq r'}^p\gamma_{rr'}x_{rj}x_{r'j}
    \label{eqn:two_factor_utility}
\end{align}
where, \\
$p$: total number of attributes\\
$x_{rj}$ : observed value of the r-th attribute for the j-th profile\\
$\beta_{r}$ : weight associated with $x_{rj}$\\
$\gamma_{r,r'}$: weight associated with $x_{rj}$ and $x_{r'j}$ \\

{\bf Modelling the Random Component:}
The multinomial logit model is based on the assumption
that the random components or errors ($\epsilon_k$ ) associated with the alternatives in a
choice set are independent and are identically distributed according to the double exponential distribution. Thus, $P(\epsilon_k<c)= \exp(-\exp(-c))$. \\

{\bf Probability of Choice}
Given the assumptions made so far, we can develop a
function to describe the probability of choice of any alternative for the individual. If
the alternative k is chosen, it implies that the random utility associated with k is the
highest (higher than the remaining item in the choice set). Let $y_j$ denote the
observed choice. Further, we let  $y_j=1$ if the j-th alternative is chosen and
0 otherwise. Then,\\
$P[y_j=1]=P[u_j>u_{j'}; j'\in S] = P[v_j+\epsilon_j>v_{j'}+\epsilon_{j'}; j'\in S]$\\
$=P[\epsilon_{j'}\leq v_j-v_{j'}+\epsilon_j]$\\
The expression can be evaluated using the assumptions made on the distributions of errors. The solution to the problem is:\\
$P[y_j=1]=\frac{1}{1+\exp(v_{j'}-v_j)}$\\
The above expression depends on unknown $\beta$ and $\gamma$ parameters. \\

{\bf Method of Estimation:}
The principle of estimation is to determine the values of parameters in the model
so as to maximize the probability (or likelihood) of the observed data. Methods like Weighted Least squares have been described in \textit{ \cite{Rao}} for estimation in such problems. We also address heterogeneity in the data by forming segments or clusters of individuals based on some background variables \textit{\cite{Virens_et_al}}.

\section{Empirical Results}
\label{model_original_sample}
A sample data of 271 individuals was collected \footnote{The survey was conducted by the students of B.Stat Third Year, 2023 of ISI.} from 3 districts of West Bengal: Kolkata, Paschim Mednipur and Barasat. Out of the total 271 individuals, 111 $(40.95\%)$ were females and 160 $(59.04\%)$ were males. Table \ref{Data-desc} shows the number of individuals at each age-group-education-gender segment. 
The design matrix as proposed in Supplementary Materials to 
\cite{GROMANN2019100136} for 11 two level attribute and profile strength being 4, was used in our study. The design matrix consisted of 120 pairs of comparison. 
In order to create a practically viable questionnaire, as a first step the odd numbered rows which represents a clear choice for a rational consumer were removed from the questionnaire. The remaining 60 questions were grouped into 5 sets of 12 questions each so that each respondent need to answer just 12 questions. 

Considering homogeneity in the market choices, the attribute weights are determined using the above proposed Multinomial Logit Model. The 11 components/indicators (\cite{INDIA_MPI}) for poverty used in this study are Nutrition (N), Child and Adolescent Mortality (CAM), Years of Schooling (YS), School Attendance (SA), Cooking Fuel (CF), Housing (H), Sanitation (S), Maternal Health (MH), Assets (A), Drinking Water (DW) and Electricity (E). 

The computations have been performed using RStudio. We first include the linear terms in the model (details are in table \ref{ROutput for linear terms}, Panel A). We observe that from the sample considered, the attributes Child \& Adolescent Mortality, School Attendance and Drinking Water are insignificant. At the next step, we fit the MNL model by removing these attributes (see table \ref{ROutput for linear terms}, Panel B). The resulting output shows \textit{statistically significantly} positive weight associated with the remaining 8 attributes, which is justifying the present model. 

\subsection{Model incorporating interactions}

Since the model (equation \ref{eqn:two_factor_utility}), also considers two factor interaction terms, the next challenge is to include the $\binom{8}{2}=28$ interactions terms (refer to Table \ref{ROutput interaction terms and Significant values}, Panel A, for detailed output).

{\bf Type of interaction between the attributes:}

We observe that there are 13 significant two-factor interaction terms present in the model. The \textit{significant} attributes can be grouped under the following 3 heads: 
\begin{enumerate}[label=(\roman*)]
    \item Health 
    \begin{itemize}
        \item Nutrition
        \item Maternal Health
    \end{itemize}
    \item Education 
    \begin{itemize}
        \item Years of Schooling
    \end{itemize}
    \item Standard of Living
    \begin{itemize}
        \item Cooking Fuel
        \item Housing
        \item Assets
        \item Sanitation
        \item Electricity
    \end{itemize}
\end{enumerate}
Using utility theory, negative estimate for interaction term indicates for the attributes being substitutes, while positive estimate indicates a complementary relation. 
The resulting model indicates following relations between the components: 
\begin{itemize}
    \item Within Heads: 
    \begin{enumerate}[label=(\roman*)]
        \item Health: \begin{itemize}
            \item Nutrition and Maternal Health are complementary. 
        \end{itemize} 
        \item Standard of Living: \begin{itemize}
            \item Housing, Electricity, Cooking Fuel, Sanitation each of these components have a substitutionary relation with Assets. 
            \item Housing and Sanitation are complementary. 
        \end{itemize}
    \end{enumerate}
    \item Between Heads: 
    \begin{enumerate}
        \item Health \& Standard of Living \begin{itemize}
            \item Nutrition is complementary to both Housing and Cooking Fuel.
            \item Nutrition is substitutionary to both Assets and Electricity. 
            \item Maternal Health and Housing are substitute to each other.
        \end{itemize}
        
        \item Standard of Living \& Education
        \begin{itemize}
            \item Assets and Years of Schooling are substitutionary to each other.
        \end{itemize}
    \end{enumerate}
\end{itemize}
\begin{figure}[h!]
    \includegraphics[width=\textwidth]{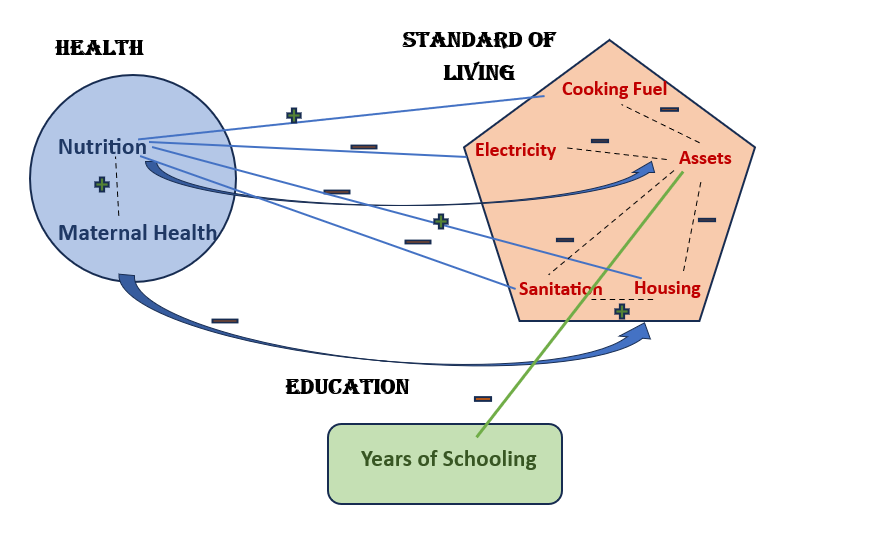}
    \caption{Figure showing the nature of interaction between the 8 attributes}
    \label{fig:web chart_interactions}
\end{figure}
\FloatBarrier

{\bf Possible Justification for these Relations:}
The Multidimensional Poverty index (\cite{OPHI}) comprises of 3 equi-weighted dimensions - Health, Education and Standard of Living. The Health dimensions consists of Nutrition, Child-Adolescent Mortality and Maternal Health with weights $\frac{1}{6}$, $\frac{1}{12}$ and $\frac{1}{12}$. The Education dimension consists of Years of Schooling and School Attendance with equal weights. The Standard of Living dimension consists of Cooking Fuel, Sanitation, Drinking Water, Electricity, Housing, Assets and Bank Account\footnote{This study does not consider the choice of Bank Account, due to ease in availability of Design matrix for 11 attributes.}, again with equal weights. The multidimensional poverty index does not consider the degree of inter-dependence between two indicators within or between the broad dimensions. Our study attempts to not only re-estimating the weights of these indicators but also determining the degree and nature of inter-relations between two attributes.  
\par As mentioned before, the multinomial model shows that some of the attributes: Child \& Adolescent Mortality, School Attendance and Drinking Water are insignificant. Further the model also shows that the coefficients associated with the attributes are different from one-another unlike the weights of the multidimensional index. The degree of 2-factor interaction further reflects the inter-relation between the attributes. 
The overall observation from the type of interaction shows that Assets have a substitutionary interaction relation with the other attributes (within head/ between heads). 
This also follows from the individual weights of the attributes with assets having the least weight. A plausible explanation can be that in times of emergencies, people would attempt to liquidate their fixed assets in order to fulfill other aspects of human conditions like housing, schooling, electricity, cooking fuel, nutrition, sanitation. Increase in demand for nourishment is supplemented by increase in intake of home-cooked meals, thus indication the complementary relation between nutrition and cooking fuel. 

\subsection{Sub-Group Level Analysis}
The output in Table \ref{ROutput interaction terms and Significant values}, Panel A, corresponds to the estimated model coefficients when the entire sample is considered. But it is an important issue to consider whether this relation is the same in different relevant subgroups of the population (e.g. according to Gender, Age, Education level etc.). This is what we attempt next. We estimate  the model for subgroups of the sample partitioned at gender, age group and education level. Education was divided into groups of below $10^{th}$ standard and $10^{th}$ pass or above. Age groups of `Below 40' and `Above or equal to 40'  were made. 
\par The comparisons of coefficients at male-female, education level and age-groups (Table \ref{ROutput interaction terms and Significant values}, Panels B \& C and all the panels of Table \ref{Tables: Education wise}) indicate a striking difference in the inter-block choices. 
\par The attributes which are significant in case of males but not in females include Years of Schooling, Assets and Electricity. Other attributes which are significant for both males and females are Nutrition, Cooking Fuel, Housing, Sanitation and Maternal Health but the coefficient estimate differ in magnitude. A possible reasoning for such differences of such choices can be that women in the areas covered in the sample are more  involved in household work and so tend to make choices of the attributes which they can directly relate to. 
Comparing the coefficients on the basis of education, it was observed that for people who received schooling below 10th standard only 3 attributes are significant: Cooking Fuel, Sanitation and Maternal Health, and for people who have atleast passes 10th all the eight attributes are significant. It may indicate that education does play a crucial role in assessing the importance of the attributes. 
Similar comparison on the basis of age shows that Cooking fuel and Assets are the only attributes significant in below 40 age group but not in other group. No suitable justification can be ascertained for age-wise difference. 

\subsection{Validity of the Results}
All the above analysis has been done under the usual distributional assumption for testing linear models. Behavioural data often behaves differently and give rise to non-standard distributions, hence invalidating the usual assumptions \textit{\cite{pihlens2004inconsistency}}. To better assess actual variability and significance of the parameter estimates, Resampling technique (Bootstrap) has been applied to estimate standard error and confidence interval of the coefficients.

The sample was grouped into age-gender-education category.  Now for each of the age-gender-education combination, the resamples of equal number (using sampling with replacement) as in the original sample were drawn and the coefficients were computed as in Section \ref{model_original_sample}. This process was repeated 10000 times and thereafter the standard error and bootstrap confidence intervals were computed. Since this process considers the various group at gender, age-group and education level it corresponds to Block bootstrap (\cite{Efron}). 
The detailed results are presented in Table \ref{Bootstrap_all_slices_result}. Following the bootstrap principle for large number of resamples, the bootstrap mean value of the coefficients must be close to the sample coefficients. However in the light of the present sample, the bootstrap mean coefficients of some linear attributes like Nutrition, Years of Schooling, Assets and Electricity differs largely intuitively. For Cooking Fuel the bootstrap $95\%$ confidence interval takes negative values which can lead to difference in interpretation.  Another major observation was that under the bootstrap method some of the interaction coefficients which were significant (under the original model) turned out to be non-significant. For instance, Nutrition \& Housing, Cooking Fuel \& Housing, Housing \& Maternal Health, Sanitation \& Assets, Sanitation \& Electricity are significant in the original model are not so in the bootstrap one while Years of Schooling \& Sanitation, Cooking Fuel \& Electricity is significant in the bootstrap model but not in the original model. 
This indicates striking differences in the estimated coefficients in the various resamples drawn, possibly due to presence of heterogeneity in the sample choices.

\section{Concluding Remarks}
The objective of this study is in applying a utility based analysis to a comparatively efficient design experiment which can capture people's perception towards the various components of a commodity. Here we studied the multi-dimensional poverty index and the relative importance of its components and their  two-factor interaction effects. Further we have discussed how to model a choice based conjoint data for determining the utility of the components and their interactions. 

Empirical results from a survey data shows the nature of  coefficients, in terms of utility derived by the individuals, their statistical significance and validity in the present framework. There has been some discrepancies in the results between the bootstrap model and the original model, which can be better understood by surveying more people, and ensuring comparative homogeneity in the data. 

This analysis can be extended to three factor interactions or more by appropriate assumptions. 
Such a design and analysis can be used in determining the nature of many-factor interaction of the many components of a commodity, which will be beneficial in decision making. But this will require a larger sample than what is available at present.

\newpage

\newpage

\section*{Appendix: Tables}

\begin{table}[h]
    \label{Gender-age-edu-femmale }
      \centering
      \caption{Description of the Data}
           \begin{tabular}{|c|c|c|c|}
            \hline
Female & Age-Group & Below 10 & $10^{th}$ Pass or more  \\\hline
&  $\leq 40$ years & 17 & 50 \\ \hline
& $> 40 $years & 20 & 24 \\ \hline 
Male & Age-Group & Below 10 & $10^{th}$ Pass or more  \\\hline
& $\leq 40$ years & 17 & 67 \\ \hline
& $> 40 $years & 18 & 58 \\ \hline 
        \end{tabular}

\label{Data-desc}
\end{table}
\begin{table}[htb!]
   \caption{Regression Output in MNL model}
   
 \label{ROutput for linear terms}
\centering

\begin{tabular}{|c | l l |l l|}  
\hline
 & \multicolumn{2}{c|}{\textit{A: All Linear Terms}} & \multicolumn{2}{c|}{\textit{B: Significant Linear Terms}} \\ \hline
\textbf{Coefficients} & \textbf{Estimate}  & \textbf{z value}  &  \textbf{Estimate}  & \textbf{z value} \\
N& 0.5691& 4.468* & 1.38508  & 16.47 *
\\
YS&  -0.3512 & -2.585*& 2.04255 & 28.52 *
\\
CF& -0.8183&  -5.096*& 1.40426  & 16.3 *
\\
H& 0.5428&  3.386 *& 1.54665  & 21.84 *
\\
S& 0.7952& 5.165*& 1.45376  & 20.23 *
\\
MH& 1.1241&  7.189*& 1.33782  & 19.4 *
\\
SA& 0.1012&  0.655&&
\\
A& 0.4337&  3.005*& 0.64526 & 11.01 *
\\
DW &  -0.1055  &   -0.722 &&  \\
E &      -0.3302  &    -2.797 * & 0.81889  & 10.48 * \\ 
CAM & -20.6751 &  -0.081 &&\\\hline

\end{tabular}

\end{table}
\FloatBarrier

\begin{table}[htb!]
\centering
\caption{Model considering the significant linear terms and their 2-factor interaction terms: Entire sample and Gender Groups}
\label{ROutput interaction terms and Significant values}
\begin{tabular}{|c | l l |l l|l l|}
\hline
 & \multicolumn{2}{c|}{\textit{A: Entire Sample}} & \multicolumn{2}{c|}{\textit{B: Females}} &  \multicolumn{2}{c|}{\textit{C: Males}} \\ \hline
\textbf{Coefficients} & \textbf{Estimate}  & \textbf{z value}  &  \textbf{Estimate}  & \textbf{z value} &  \textbf{Estimate}  & \textbf{z value} \\
N & 1.24043  & 5.339 * & 1.50697  & 12.448 *& 1.25048  & 4.004 *\\
YS & 3.53095  & 18.95  * && & 3.50011  & 14.538 *\\
CF& 1.01376 & 6.214 * & 0.68302 & 5.301 *& 1.09868  & 5.07 *\\
H & 1.73409  & 9.76 * & 1.26764 & 7.036 *& 1.68098 & 7.283 * \\
S & 1.9438  & 12.926 * & 1.02929 & 7.494 *& 1.95179  & 9.893 *\\
MH & 1.74674  & 14.094 * & 0.99494 &  8.444 *& 1.68354 &  10.349 *\\
A& 2.12305  & 16.83 * &&& 2.0753 &  12.554 * \\
E & 1.36538 & 8.189 * &&& 1.24391 &  5.637 *\\
N*YS & -0.46711  & -1.631 && & -0.43686  & 0.25016  \\
  N*CF & 2.14002  & 6.107 * & 0.06901 &  0.247& 2.24966 &  4.798 *\\
  N*H   & 1.63791 & 4.625 *  & -0.19596  & -0.76& 1.83511 &  3.904 * \\
  N*S   & -0.81814  & -3.274 * & 0.04084 & 0.172 & -0.97876 &  -2.873 *\\
  N*MH   & 2.28394 & 6.568 * & 2.46945 & 6.617& 2.26387 &  4.952 *\\
  N*A & -0.97049  & -4.153 * &&& -0.93355 &  -2.956 *\\
  N*E   & -1.93248  & -7.354 * &&& -1.51217 &  -4.429 *\\
  YS*CF & 0.34302  & 0.882  && & 0.22965 &  0.451 \\
  YS*H   & 0.11854  & 0.209   &&& 0.20664 &  0.272\\
  YS*S   & 13.65828  & 0.024    &&& 13.68814 &  0.019\\
  YS*MH   & -0.19958  & -0.714 &&& 0.28876 &  0.769\\
  YS*A & -2.9624  & -9.668 *&&& -2.90671 &  -7.183 \\
  YS*E   & 16.69124  & 0.065 &&& 16.60703 &  0.049 \\
CF*H   & -0.0963 & -0.394  & -0.04535  & -0.188& -0.12047 & -0.392\\
CF*S   & 0.12305  & 0.447     & 0.08378 &0.298& 0.11645 &  0.336 \\
CF*MH   & 0.86946  & 3.39 * & 0.49612 & 1.67 & 0.66565 &  2.07 *\\
CF*A & -1.76139 & -6.18 * &&& -2.10907 &  -5.662 * \\
CF*E   & 18.60707  & 0.072 &&  & 18.44663 &  0.054\\
  H*S   & 1.74064  & 6.214 * & 1.24733  & 3.379& 1.79359 &  4.856 * \\
  H*MH   & -0.52949  & -2.148 *& 0.24754  & 0.802 & -0.40203 &-1.257 \\
  H*A & -2.33545  & -11.276 * &&& -2.21382 & -8.325 *\\
  H*E   & -0.2566  & -0.664  && & -0.02593 &-0.051\\
  S*MH   & -0.464  & -1.743  & -1.85661  & -7.528 & -0.11987 & -0.338\\
  S*A & -0.61316  & -2.718 * && & -0.17116 &-0.546\\
  S*E   & -0.97918 & -3.117 * && & -0.83449 & -2.03 *\\
  MH*A & -0.04204  & -0.203   &&& -0.16667 &  -0.613 \\
  MH*E   & -1.3609 & -5.794 * &&& -1.51301 &  -4.94 *\\
  A*E   & -2.62369 & -10.684 * &&& -2.68925 &  -8.244 *\\
\hline
\end{tabular}
\end{table}
\FloatBarrier

\hoffset -1.5cm
\begin{table}[htb!]
\centering
\caption{Regression Output for Education and Age Groups}
\label{Tables: Education wise}
\begin{tabular}{|c  |l l |l l|ll|ll|}  
\hline
 & \multicolumn{2}{c|}{\textit{A: Below 10th Standard}} & \multicolumn{2}{c|}{\textit{B: 10th Pass or More}} &  \multicolumn{2}{c|}{\textit{C: Age $\leq 40$}} & \multicolumn{2}{c|}{\textit{D: Age $>40$}}\\ \hline
\textbf{Coefficients} & \textbf{Estimate}  & \textbf{z value}  &  \textbf{Estimate}  & \textbf{z value}  &  \textbf{Estimate}  &\textbf{z value}  
 & \textbf{Estimate}  &\textbf{z value} 
\\
N & && 1.49643  & 5.581 *  &  1.82885  &14.817 *
 & 1.53103  &7.711 * 
\\
YS & &&3.56569  & 15.953 *  &  &
 & 2.10065 &9.902 * 
\\
CF& 0.3077  & 2.616 *& 0.99315  & 5.323 *  &  1.10666  &8.709 *
 & -0.02044&-0.114  
\\
H & &&1.91034  & 9.3 *  &  1.28917 &7.344 *
 & 1.67813 &7.307 * 
\\
S & 0.957  & 8.267 *&1.99094  & 11.321 *  &  0.93689  &6.436 *
 & 1.92354 &10.401 * 
\\
MH &1.04187 & 8.557 * & 1.7336 & 12.083 *  &  1.10074  &9.929 * 
 & 1.17703  &7.484 * 
\\
A& &&2.08785  & 14.353 *  &   1.09998  &10.324 *&&

\\
E & &&1.32928  & 6.935 *  & 
 & &   0.2606  &1.307  
\\
 N*YS  & &&-0.6809  & -2.071 *  &  &
 & -0.63166  &-1.693 
\\
 N*CF & &&2.00948 & 5.08 *  &  -0.14533  &-0.548
 & 1.72249 &4.247 * 
\\
 N*H  & &&1.30531 & 3.246  *  &  -0.28331  &-1.228
 & 0.63583  &1.531   
\\
 N*S  & &&-1.00841 & -3.468  *  &  -0.07304  &-0.355
 & -1.06645  &-3.434 * 
\\
 N*MH  &&& 2.34642  & 5.845 *  &  2.84496  &7.547 * 
 & 1.87659  &4.48 * 
\\
 N*A & &&-1.09075  & -3.99 *  &  -1.27695  &-5.768 *  &&
\\
 N*E  & &&-1.78711  & -5.977 *  & 
 &  & -0.98654  &-2.649 * 
\\
 YS*CF & &&0.16986  & 0.375   &  &
 & 1.47538  &2.582 * 
\\
 YS*H  & &&0.15108 & 0.234    &  &
 & 0.42358  &0.535   
\\
 YS*S  & &&13.96719  & 0.021    &  &
 & 14.4896  &0.03   
\\
 YS*MH  & &&-0.42431  & -1.332     &  &
 & 0.47271  &1.375   
\\
 YS*A & &&-2.69289  & -7.534 *  &  &
 &&
\\
YS*E  & &&16.64808  & 0.056    &   &
  & 16.38397  &0.064  \\
CF*H  & &&-0.17676  & -0.602     &  -0.20758& -0.945& 0.28165& 0.881\\
CF*S  & -0.4242  & -1.514  & 0.26768  & 0.794     &  -0.20083& -0.777& 0.60254& 1.627 
\\
CF*MH  & -0.4272  & -1.815  &1.20123  & 3.813  *  &-0.07901& -0.289& 1.54031 &4.743 *   
\\
CF*A &&& -1.77394 & -5.337 *  & -1.21433 &-4.029 *
 & &
\\
CF*E  & &&18.68235  & 0.063     & &&17.74389 &0.069
\\
 H*S  & &&1.50992 & 4.72 *  &2.44161& 5.981 * &0.796& 2.139 * 
\\
 H*MH  & &&-0.66324  & -2.313 *  & 0.58094& 1.798& -0.14011 &-0.453
\\
 H*A & &&-2.43822  & -10.057 *  &  -1.50857 &-6.437 *
 & &
\\
 H*E  & &&-0.38539  & -0.872    &  & &0.934 &1.757  
\\
 S*MH  &-0.5886  & -2.897 * & -0.68783 & -2.318 *  & -2.26645 &-9.51 * &-0.34916 &-1.073 
\\
 S*A & &&-0.56558 & -2.173 *  &  -0.33688  &-1.495 
 & &
\\
 S*E  & &&-0.9396  & -2.572 *  &  &
 & 0.25614 &0.578 
\\
 MH*A & &&-0.01946  & -0.081    &  -0.2578  &-0.928  & &\\
 MH*E  & &&-1.57662  & -5.815 *  &  && -0.45704 &-1.397\\
 A*E  & &&-2.50188  & -8.898 *  &  & & &\\

\hline

\end{tabular}
\end{table}
 \FloatBarrier

\hoffset 0cm
\begin{table}[h]
\caption{Bootstrapping Results (* significance according to Bootstrap)}
    \label{Bootstrap_all_slices_result}  
\begin{tabular}{|l l l l l l l|}       
\hline
\textbf{Terms} & \textbf{Observed} & \textbf{Bootstrap} & \textbf{Bootstrap} & \textbf{Z-value} & \textbf{P value} & \textbf{Confidence} \\
& \textbf{Coefficient} & \textbf{mean values} & \textbf{SE} & &  & \textbf{Interval} \\
\hline
N & 1.24* & 1.41 & 0.41 & 3.06 & 0.0022 & ( 0.22 , 1.71 )* \\
YS & 3.53* & 2.60 & 1.34 & 2.64 & 0.0084 & ( 3.23 , 7.06 )* \\
CF& 1.01* & 1.03 & 0.46 & 2.22 & 0.0267 & ( -0.11 , 1.87 )* \\
H & 1.73* & 1.61 & 0.41 & 4.22 & 0.0000 & ( 1.34 , 3.46 )* \\
S & 1.94* & 1.77 & 0.45 & 4.35 & 0.0000 & ( 1.3 , 3.13 )* \\
MH & 1.75* & 1.65 & 0.42 & 4.12 & 0.0000 & ( 0.66 , 2.52 )* \\
A& 2.12* & 1.77 & 0.66 & 3.22 & 0.0013 & ( 1.56 , 4.24 )* \\
E & 1.37* & 1.03 & 0.64 & 2.14 & 0.0321 & ( 0.5 , 2.74 )* \\
 N*YS  & -0.47 & -0.23 & 0.48 & -0.98 & 0.3262 & ( -1.88 , 0.08 ) \\
 N*CF & 2.14* & 1.54 & 0.94 & 2.28 & 0.0224 & ( 1.52 , 4.62 )* \\
 N*H  & 1.64* & 1.14 & 0.97 & 1.68 & 0.0925 & ( 1.09 , 4.32 ) \\
 N*S  & -0.82* & -0.66 & 0.37 & -2.20 & 0.0279 & ( -1.73 , -0.36 )* \\
 N*MH  & 2.28* & 2.05 & 0.71 & 3.21 & 0.0013 & ( 1.69 , 4.78 )* \\
 N*A & -0.97* & -1.00 & 0.48 & -2.03 & 0.0427 & ( -1.94 , 0.06 )* \\
 N*E  & -1.93* & -1.40 & 0.90 & -2.14 & 0.0323 & ( -3.86 , -1.05 )* \\
 YS*CF & 0.34 & 0.34 & 0.49 & 0.70 & 0.4812 & ( -0.76 , 1.27 ) \\
 YS*H  & 0.12 & 0.07 & 0.55 & 0.22 & 0.8290 & ( -1.05 , 1.18 ) \\
 YS*S  & 13.66 & 11.36 & 5.11 & 2.67 & 0.0075 & ( 11.6 , 27.32 )* \\
 YS*MH  & -0.20 & 0.02 & 0.46 & -0.44 & 0.6627 & ( -1.55 , 0.33 ) \\
 YS*A & -2.96* & -2.11 & 1.29 & -2.30 & 0.0215 & ( -5.92 , -2.44 )* \\
 YS*E  & 16.69 & 12.20 & 7.50 & 2.23 & 0.0260 & ( 15.4 , 33.38 )* \\
CF*H  & -0.10* & -0.12 & 0.45 & -0.22 & 0.8297 & ( -1.02 , 0.98 ) \\
CF*S  & 0.12 & 0.10 & 0.30 & 0.40 & 0.6857 & ( -0.4 , 0.87 ) \\
CF*MH  & 0.87 & 0.70 & 0.47 & 1.84 & 0.0656 & ( 0.26 , 2.03 ) \\
CF*A & -1.76* & -1.53 & 0.76 & -2.30 & 0.0212 & ( -3.52 , -0.79 )* \\
CF*E  & 18.61 & 13.51 & 8.09 & 2.30 & 0.0215 & ( 17.93 , 37.22 )* \\
 H*S  & 1.74*& 1.61 & 0.52 & 3.35 & 0.0008 & ( 1.13 , 3.48 )* \\
 H*MH  & -0.53* & -0.15 & 0.76 & -0.69 & 0.4876 & ( -3.07 , 0.05 ) \\
 H*A & -2.34* & -1.89 & 0.87 & -2.67 & 0.0075 & ( -4.68 , -1.92 )* \\
 H*E  & -0.26 & 0.01 & 0.74 & -0.35 & 0.7291 & ( -2.83 , 0.51 ) \\
 S*MH  & -0.46 & -0.61 & 0.60 & -0.77 & 0.4425 & ( -1.18 , 1.15 ) \\
 S*A & -0.61* & -0.39 & 0.36 & -1.69 & 0.0908 & ( -1.71 , -0.27 ) \\
 S*E  & -0.98* & -0.67 & 0.74 & -1.33 & 0.1837 & ( -3.1 , -0.02 ) \\
 MH*A & -0.04 & -0.15 & 0.47 & -0.09 & 0.9284 & ( -0.53 , 1.11 ) \\
 MH*E  & -1.36* & -0.11 & 0.62 & -2.18 & 0.0291 & ( -2.72 , -0.83 )* \\
 A*E  & -2.62* & -0.98 & 1.18 & -2.22 & 0.0267 & ( -5.24 , -2.03 )* \\
\hline
\end{tabular}
\end{table}

\end{document}